# An Empirical Evaluation of Preconditioning Data for Accelerating Convex Hull Computations

**José O. Cadenas**[1], **Graham Megson**[2]

**1** School of Systems Engineering, University of Reading, Reading UK, **2** University of Westminster, London, UK

## Abstract

The convex hull describes the extent or shape of a set of data and is used ubiquitously in computational geometry. Common algorithms to construct the convex hull on a finite set of $n$ points $(x,y)$ range from $O(n\log n)$ time to $O(n)$ time. However, it is often the case that a heuristic procedure is applied to reduce the original set of $n$ points to a set of $s < n$ points which contains the hull and so accelerates the final hull finding procedure. We present an algorithm to precondition data before building a 2D convex hull with integer coordinates, with three distinct advantages. First, for all practical purposes, it is linear; second, no explicit sorting of data is required and third, the reduced set of $s$ points is constructed such that it forms an ordered set that can be directly pipelined into an $O(n)$ time convex hull algorithm. Under these criteria a fast (or $O(n)$) pre-conditioner in principle creates a fast convex hull (approximately $O(n)$) for an arbitrary set of points. The paper empirically evaluates and quantifies the acceleration generated by the method against the most common convex hull algorithms. An extra acceleration of at least four times when compared to previous existing preconditioning methods is found from experiments on a dataset.

## Introduction

Computing the convex hull on a set of $n$ 2D points is a first pre-processing step to many geometric algorithms and in practical applications (e.g. computer visualization, maps, rover path finding and home range [1]). Indeed, one can say with confidence, that finding the boundary of a given set of points is a fundamental problem in providing fast algorithms in many modern day mobile devices, games consoles, digital cameras, and client-server (web) applications that seek to reduce and create knowledge or patterns from raw data collection. The convex hull is used to analyze multidimensional random data from time series [2] now typically associated to big data sets due to data collected from sensors attached to billions of interconnected smart devices.

Most known convex hull algorithms are of time complexity $O(n\log n)$ [3]; these methods are general in the sense that they do not impose any restriction in the order in which points are considered. Linear complexity ($O(n)$) methods do exist but require a set of points that are ordered in some way, for example [4] requires an order where the points form a simple polygonal chain. Such orderings are not always easy to produce given the process of data collection.

Regardless of the time complexity of an algorithm, reducing the set of $n$ points down to a set of $s \leq n$ points would result in faster computations, provided that the smaller set preserves the convex hull of the original (bigger) set, and provided that the time taken to perform the reduction offsets the cost of processing $n$-$s$ points in a convex hull building method. This reduction is often used as the first step in implementation of convex hull algorithms to improve their performance [5]. Here, a new and surprisingly simple approach to perform a reduction is empirically analyzed [6], exhibiting three distinct advantages over other methods.

— First, we show that, for all practical cases, the method is linear and can be applied before any 2D known convex hull algorithms..
— Second, no explicit sort of points is required.
— Third, by construction the reduced set of data forms a simple polygonal chain and hence straightforwardly prepares the data for linear methods.

The second point is significant because typically reduction methods in the existing literature require an explicit sort of the points along a particular direction [5].

We show through experimental evaluation that the method makes faster convex hull computations for both linear and non-linear algorithms. The analysis is applied to a data set derived from 3D scans of animal skeletons [7]; this dataset is similar in structure to many common 3D data used in neuroscience that are prime candidates for big data sets and analysis. A comparison is given of the achieved reduction of the proposed method against two other key preconditioning methods taken from the literature. The final convex hull derived from the reduced set of points is then computed using the most common convex hull algorithms specifically Graham Scan, QuickHull, and Jarvis March [8] to benchmark the speedup benefit gained. Since our approach readily produces a polygonal chain, for completeness, we also apply it to the method of Melkman which is not possible with the other preconditioning methods.

In the sequel, and to help clarify the discussion, note that the reduced set of (preconditioned) points is contained in an array of size $p$; the array may contain entries that do not correspond to any of the original points. Fast mechanisms to skip over these invalid entries are also discussed and empirically analyzed where it is shown that for all practical purposes the method remains linear in time.

## Methods for preconditioning data

### Proposed heuristic

Assume a 2D square of side $p$, with integer points whose coordinates are in the range 1, …, $p$ from which it follows that $1 \leq x, y \leq p$. As a small example consider the set of $(x, y)$ points on the left of Fig. 1, given in any order as an array $P$.

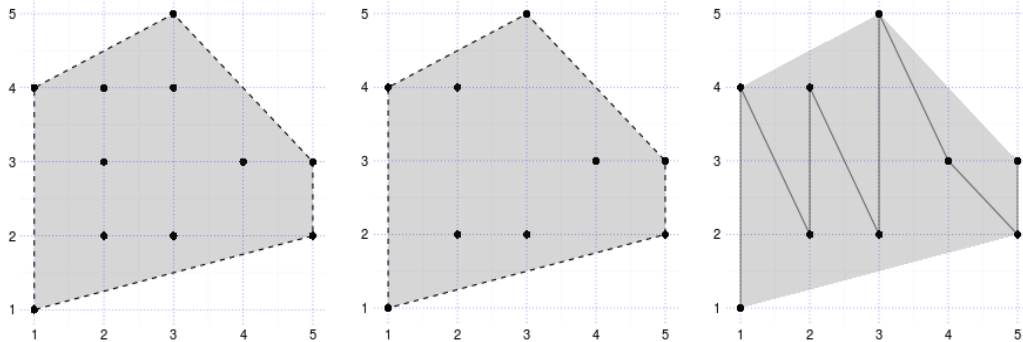

Fig. 1. Left: $(x, y)$ integer points on a 2D grid with $p$ = 5. Center: Points with minimum and maximum $y$ values for each $x$ coordinate. Right: A polyline.

Assume an array $L$ of $p$ elements with each element $L[i]$, $i$ = 1, …, $p$, initialized to $(p+1, -1)$, so $L = [(p+1,-1), …, (p+1, -1)]$. Consider the following pseudo-code:

**ALGORITHM 1.** Reduction of Points

**Input:** An array $P$ of $n$ points with $(x,y)$ integer coordinates
**Output:** An array $L$ of $p$ entries, with each entry either $(ly, hy)$ or $(p+1, -1)$
$L = [(p+1, -1), …, (p+1, -1)]$

**foreach** point **in** $P$ **do**
    $x_i, y_i$ = point
    $y_v, y_u = L[x_i]$
    $ly = \min(y_i, y_v)$
    $hy = \max(y_i, y_u)$
    $L[x_i] = (ly, hy)$
**end**

After all $n$ points of $P$ have been processed by the above routine, $L = [(1,4), (2, 4), (2,5), (3,3), (2,3)]$. $L[1] = (1, 4)$ since, $y = 1$ is the minimum point (min); and $y = 4$ is the maximum (max) point for column $x = 1$. This reduced set is shown in the center of Fig. 1. Intuitively the convex hull on the left is the same as the convex hull on the center. Since local convexities of the boundary points are conserved and collinear points are removed we need only to consider the min and max points on an $x$ (or $y$) file when deriving a convex hull [3]. This observation in itself is trivial but when combined with a suitable hashing technique creates a very fast preconditioning.

*Building a polyline.* Scanning $L$ along $x$ builds a simple polygonal chain, since joining all points of the reduced set $s \leq n$ creates edges that do not intersect. For each valid point in $L$ (one different to $(p+1, -1)$) joining $ly$ to $hy$ (min to max), and then from $hy$ (max) to $ly$ (min) of the next valid point, forms a simple polygonal chain. We formalize this principle with the following lemma:

**Lemma 1**: For an array $L$ with entries $(ly_i, hy_i)$ $i = 1, …, p$ so that $1 \leq ly_i, hy_i \geq q$ are the minimum and maximum points respectively at X coordinate $i$, found inside a bounding box of $m = p \times q$ points with integer coordinates, there exists a simple chain joining all the minimum and maximum points.
**Proof**: [by construction]
The chain is formed by scanning entries of $L$ along $i = 1, …, p$. Each entry $i$ of $L$ contains one or two points or no points of the bounding box. Let's refer to an entry $(ly_i, hy_i)$ of $L$ such that $1 \leq ly_i, hy_i \geq q$ as corresponding to valid points. An entry $(ly_i, hy_i)$ where $q \leq ly_i, hy_i \leq 1$ corresponds to a column in the bounding box with no valid points. An entry $i$ in $L$ with no valid points is skipped. If the entry $i$ has a single valid point $(ly_i = hy_i)$ that point is kept. For two valid points in an entry $i$, the two valid points are joined with an edge that runs from $(i,ly)$ to $(i,hy)$. The whole chain is formed by connecting edges in an entry $i$ (or single point in entry $i$ or skipping entry $i$) to edges in adjacent entries (or single point in adjacent entry or skipping adjacent entry) until all valid points are connected. Adjacent entries $i$ and $i+1$ are connected by an edge from $(i, hy)$ to $(i+1, ly)$ or from $(i, hy)$ to $(i+k, ly)$ where entries $i+1, …, k-1$ had no valid points. This creates a simple chain covering all the valid points in $L$ since no edges intersect.

Lemma 1 is not the only way to form such a chain we can use a $y$-scan or a clockwise or anticlockwise traversal of the points but Lemma 1 accurately described the points contained in array $L$ after a scan using Algorithm 1 (see Fig. 1 on the right).

*Quick analysis.* The routine of Algorithm 1 above visits each point of *P* once, therefore *L* is built in O(*n*) time. Scanning *L* to build the polygonal chain, Lemma 1, then takes O(*p*) time and provided that $p \leq n$ the whole method of building the polygonal chain takes O(*n*) time. As up to $n = p^2$ points may be in the original set, a density of points $s/n > 2/p$ makes the method O(*n*). If $p > n$, then there is a simple method to skip unused entries in *p* which remains fast – however for our data set the above holds and in practical situations with relatively dense data this condition usually applies. A simple analysis shows that the method potentially reduces the percentage of original points to $1 - 2/p$.

*Correctness of Algorithm 1:* See appendix.

Below we consider some existing methods that also precondition the points, as you will see they are somewhat more complex than the current method.

**Common heuristic to reduce a set of 2D points**

A simple algorithm to compute the convex hull on a set of 2D points is presented in [5]. The algorithm has a worst case time complexity of O(*n*log*n*). It is composed of three steps, with a first step being a preprocessing procedure that finds convex points from which a preliminary reduction of points can be made before attempting to build the whole convex hull. The preprocessing follows by determining the minimum and maximum along *x* and *y* coordinates. With these points, four or fewer external regions are formed. All the points found inside these external regions are discarded since they cannot belong to the convex hull. Applying this heuristic to the small example on the left of Fig. 1, three external regions are formed (outside the convex hull and inside the box). All points internal to the pentagon (convex hull of the figure, shaded) can be discarded. In this trivial example, this heuristic has not left any points to be further considered. In general, it is reported that for large values of *n* the number of points is reduced to less than half. For simplicity we refer to this method as the AT method.

**An iterative heuristic to reduce a set of 2D points**

This heuristic eliminates all interior points to a pseudo-hull that is built over several iterations [9]. The process starts by applying the AT method to build an initial 2D polygon. Interior points to this polygon are discarded. A new iteration begins. For each edge of this initial 2D polygon the furthest point (external to the polygon) is computed and then the edge is replaced by connecting this furthest point to the two vertices of the edge. The polygon gets extended with new edges and points internal to this expanded polygon are discarded. This process is repeated a number of times; it is reported that most of the interior points are discarded after four iterations. This method is referred here as the TZTM method.

**A real-world dataset**

For comparison we employ a dataset of 3D models generated from a 3D scanner of mammals' skeletons [7]. It is composed of 14 animals. These types of data sets are typical of the information stored as a result of big data sets from sensing or simulation of physical processes. A 2D projection on a plane with integer coordinates was performed resulting in the number of points as shown comparatively in Fig 2 along with density of points calculated as $n/(p*q)$; (bigger area for a mammal implies bigger number of points for that mammal in the dataset).

Empirical evaluation of preconditioning for fast convex hull computation

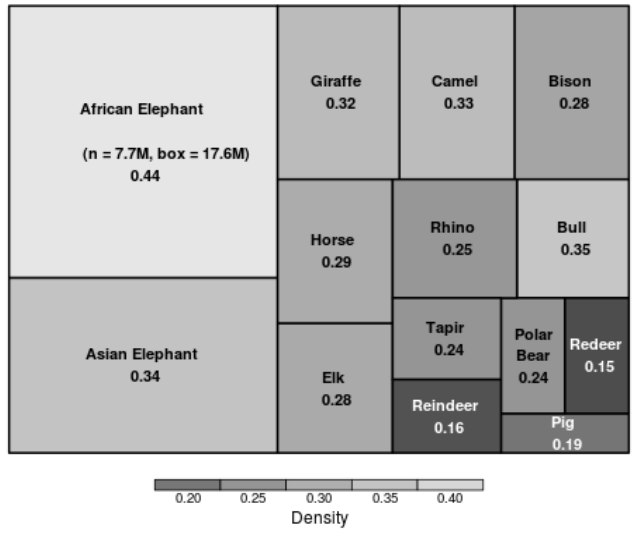

Fig. 2. Relative number of points for each mammal in the dataset. African Elephant has $n = 7.7 \times 10^6$ points in a box $p*q$ with $p = 7129$, $q = 2475$ for a size of $17 \times 10^6$ and density 0.44. The density is $n$/size.

The two alternative methods described above as AT and TZTM were applied to this dataset. The results are shown in Fig. 3; the reduction is given as a percentage, 0.0 would indicate all $n$ points were reduced to $s = 0$. Notice that for TZTM method, the reduction after 2, 3, and 4 iterations are presented. The proposed method obtains a percentage of reduction of points of over 98%. AT method gives a percentage of reduction of points between 83%-91%. The iterative heuristic applied in TZTM will gradually discard more points, but requires usually three or more iterations before it gets better than the proposed method here (and even so for Asian Elephant, this is still not the case). An interesting point to notice is that, the reduction of points achieved by the proposed method seems to be independent of the shape of the animals; this is appreciated if the results are plotted as a boxplot. In a boxplot the two elephants are outliers and the reduction is more effective than the other mammals in the 3rd and 4th iteration for TZTM method probably due to the squareness appearance of the 2D projection used here for those two mammals as compared to the rest of mammals.

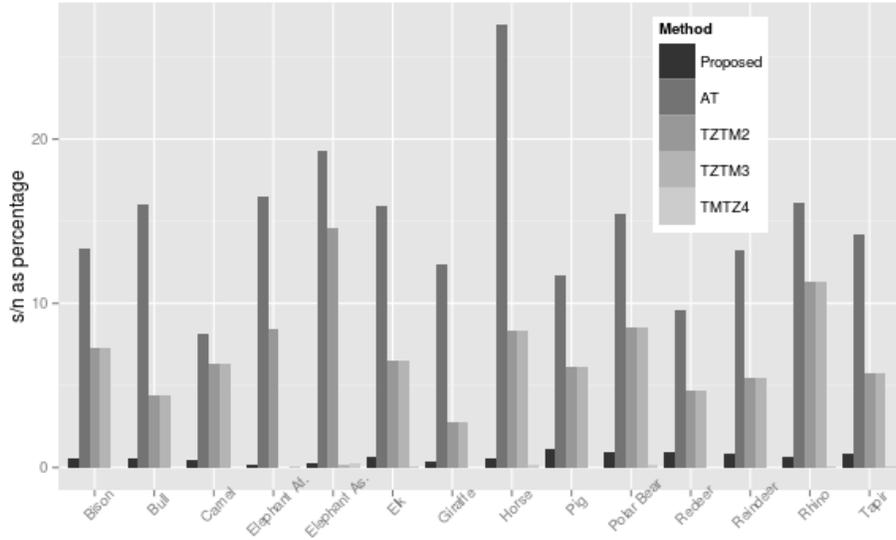

Fig. 3. Reduction of data points by preconditioning using the Proposed, AT, and TZTM methods, (2, 3 and 4 iterations for the TZTM method).

**Implementation of the proposed heuristic**

In a general case, a method to reduce a set of n points to a set of s points, before building a convex hull is given by:

**Step 1**: Find the maximum and minimum $x$ and $y$ in the point set $P$ (define $p$, $q$).
**Step 2**: Translate the point set $P$ into a point set $P'$ using $(x', y') = (x$-$p$+1, $y$-$q$+1).
**Step 3**: Build array $L[i]$ from the set of points $P'$ by applying the routine given by Algorithm 1.

Step 1 and 2 are both O($n$). As from step 1 $p$ is known, step 3 reduces set $P$ to a set $P'$ of $s \leq n$ and is O($n$) for $p \leq n$; it will be seen later that for practical cases the method can be considered O($n$) when $p > n$. The whole procedure of steps 1 through 3 is O($n$) for all practical cases. From Fig. 4, it is clearly seen that $s < n$ in the dataset used in this paper (also $s \leq 2p$).

Notice performing a sweep of points along the $min(p, q)$ ($p$, $q$ obtained from step 1) allows for the greatest reduction (instead of using p as Algorithm 1). For the mammals' dataset, Fig. 4 shows $p$, $q$ values, number of points remaining after a scan along $p$ (labelled $rx$) and also the further reduction by a second scan along $q$ (labelled $rxy$).

In many problems, the size of $p$, $q$ is already known (e.g. detecting the boundary of binary images, collision detection, and cloud segmentation of a geographical area) and so the size of $L$ is pre-computed without a need for step 1. In any case step 1 is also required for the common AT heuristic and consequently also in the TZTM method. So, steps 2 and 3 of the proposed method are the most relevant for a fair comparison against the alternative two methods we have considered.

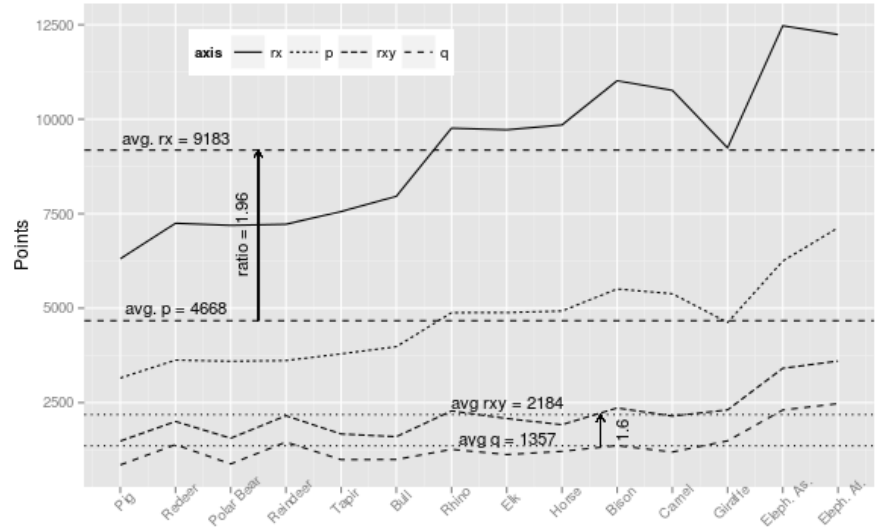

Fig. 4. Reduction of data points by Algorithm 1 on a mammals' dataset. After a scan of $n$ points along $x$ axis (size $p$) up to $rx = 2p$ points remain. A further scan on $rx$ points along $y$ axis (size $q$) up to $rxy = 2q$ points remain.

The AT method uses a cross-product (2 multiplications and 5 additions); TZTM uses an additional distance calculation to discard points while the method here proposed uses just comparisons and insertions into arrays. Consequently the cost of each processing step of the proposed method is also fast compared to these two existing schemes.

The major difference between our approach and the AT, TMTZ approaches is that building array $L$, in step 3, can be thought of as a process of including points that are candidates to belong to the final convex hull rather than a process of discarding points that cannot be in the convex hull. Also, notice that step 3 does not require sorting the points – it is a simple hash function (or lookup operation) on the axis.

## Results

Clock time has been used during all the experiments using a timer with granularity in the order of microseconds. For each dataset, each different experiment was repeated one hundred times and the average time across all runs is reported to allow for system processes.

**Impact on reduction time for building the convex hull by the proposed heuristic**

For step 1 it is not necessary to compute $y_{min}$ nor $y_{max}$ if output array $L$ will be built after a scan along $x$ only (so only $x_{min}$ and $x_{max}$ are computed in step 1). The total execution time is given in Fig. 5, in milliseconds along with the total execution time (including the time to compute $x_{min}$, $x_{max}$, $y_{min}$, and $y_{max}$) for the reduction method referred as AT. For the TZTM method, executed time is shown for 2, 3, and 4 iterations. As the code for these are not available, they have been coded for this work to the best of our ability based from the descriptions in the original papers.

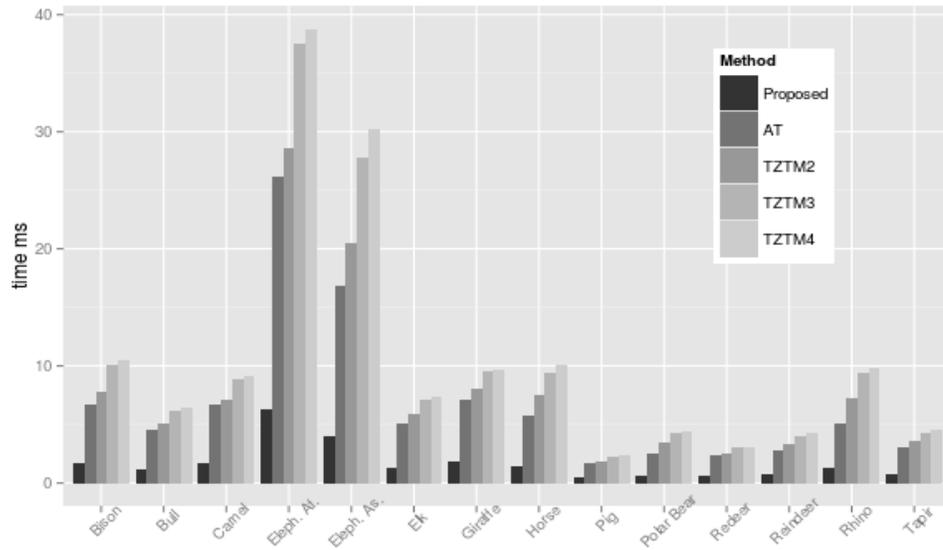

Fig. 5. Execution time spent to perform a reduction of points on a mammals' dataset by three methods.

Fig. 6 shows the split of execution time across the three different steps. In order to allow for a more direct comparison to the AT reduction method steps 2 and 3 were merged into a single step. Step 1 finds $x_{min}$ and $x_{max}$ across all points, while steps 2-3 perform the actual reduction. Similarly, step 1 under AT is the time spent to find $x_{min}$, $x_{max}$, $y_{min}$, and $y_{max}$ for all points, while step 2 is the actual procedure to reduce the points. The proposed method shows a linear trend in its measured execution time of $8.15 \times 10^{-7} n$ seconds. For comparison, AT method also shows a linear trend in its measured execution time of $3.42 \times 10^{-6} n$ seconds (note, the reduction of points using AT is linear, not building the final hull). From here, a factor of 4 is observed favoring the proposed method against AT (see numbers on top of bars on Proposed method).

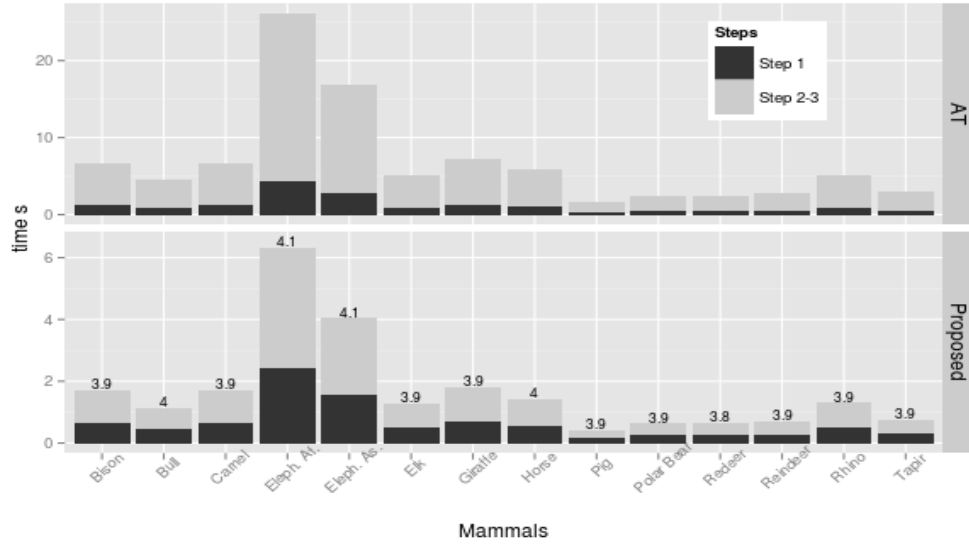

Fig. 6. Time spent in internal steps by AT and Proposed methods. Note the scale on the time axis for each method. The ratio of execution time for AT to execution time for Proposed is given on top of the bars.

**Proposed heuristic impact on the convex hull computation**

In this section, the execution time using the proposed reduction methods is considered. First, the time to compute the convex hull on the original $n$ number of points is computed for a given algorithm and reported as $t_n$. The total execution time to compute the convex hull on the reduced set of $s$ points is then given by the time it takes to perform the reduction, $t_r$, plus the time it takes to compute the convex hull on the reduced set, $t_s$. We have chosen three well known convex algorithms mostly taking by popularity of citations in different fields.

*Using QuickHull.* Recall the proposed heuristic (Section 2.5) applies the method by a scan along the $x$ coordinate and a second scan along the $y$ coordinate may follow for further reduction. Fig. 7 annotates the overall time to compute the convex hull on a reduction along the $x$ coordinate as $t_{rx}$, while $t_{rxy}$ is the time it took when the convex hull was computed by a reduction along $x$ and $y$ coordinate. QuickHull [10] is used to build convex hull either for $n$ points ($t_n$) or the reduced set of points. Fig. 7 shows that the extra benefit of having a reduction along the $y$ coordinate after a scan along the $x$ coordinate is marginal and thus is not pursued any further in this paper. The figure annotates the percentage of time to compute the final convex hull on the reduced set of points; building the hull takes under 5% of the total execution time. It also annotates the speed up in execution time observed across the dataset. It is in the range of 2.8-3.4. Note that the figure says that to compute the convex hull on n points (for the used dataset) it is faster, first to reduce the points and then compute the hull on the reduced set of points, rather than computing the hull directly on the n points. It does not make any comparison judgment on the QuickHull algorithm per se.

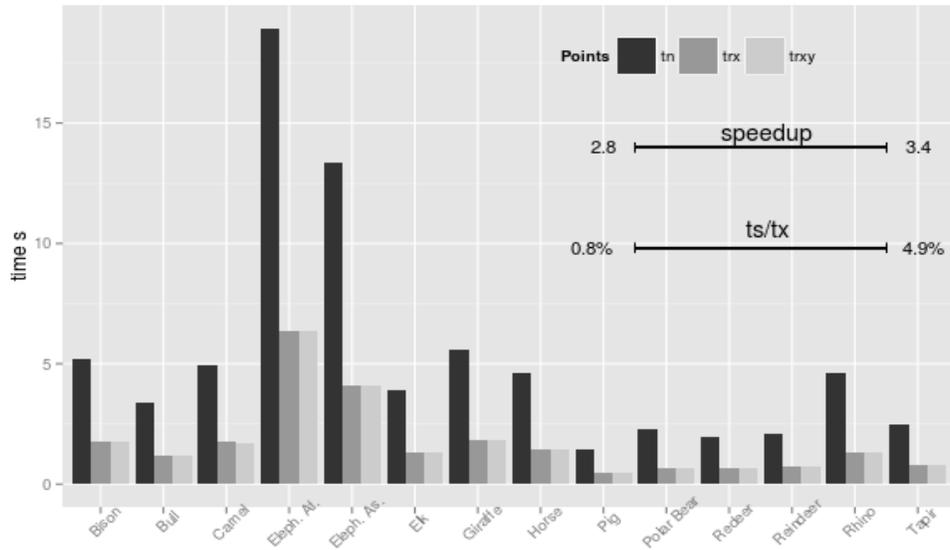

Fig. 7. The $n$ points of each mammal is reduced by the proposed method to $rx$ points, and the time to build the hull using QuickHull algorithm $trx$, ($trxy$ is extra time to further reduce along $y$ axis as well) is compared against the time to compute the hull on $n$ points by Quickhull ($tn$).

*Using Graham Scan.* The impact of the proposed preconditioning method in the case when the Graham Scan algorithm [11] is used to compute the hull is seen in Fig. 8. The amount of acceleration provided by the reduction for this dataset is of around 5 from as observed from the figure. Hidden from the figure (and from analysis of the tabulated results) the time spent doing the reduction can be as high as 99% of the total time and is consistently over 90% across the dataset.

*Using Jarvis March.* The impact of the proposed preconditioning method in the case when the Jarvis March algorithm [12] is used to compute the hull is seen in Fig. 8. The amount of acceleration seen for this dataset is greater than 15 and under 50. The speed up was calculated as the time taken to compute the hull on n points using Jarvis March to the time to reduce n points by the proposed method and then computing the final hull using Jarvis March. Also, not appreciated from the figure, the time spent doing the reduction can be as high as 90% of the total time (this was the case for the two elephant animals in the dataset).

Figure 8 shows that the use of the proposed method is useful to accelerate the convex hull computation regardless of the algorithm used to build the hull. Rather than applying a given convex hull algorithm on the original set of $n$ points, it is worth spending time in reducing the set first before giving a reduced set of points to the algorithm. As the method to reduce is simple and fast, most of the time to build the final hull is now spent in performing the reduction. The amount of acceleration depends on the algorithm as expected.

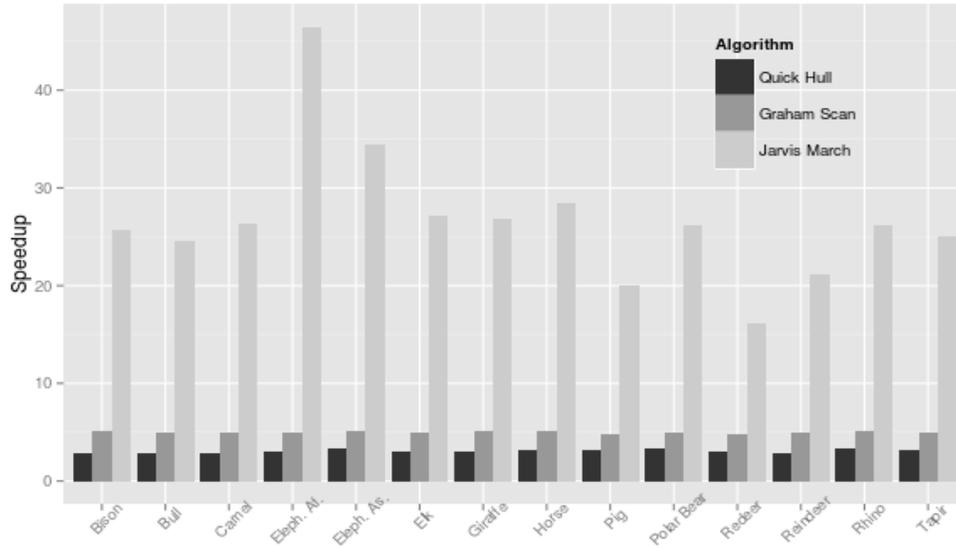

Fig. 8. Speedup gained in execution time to compute the convex hull on the set of mammal dataset by first reducing points by the proposed method and then computing the hull on the reduced.

**Impact on the convex hull computation by AT and TZTM methods**

For comparative purposes, this section presents overall execution time for computing the convex hull for the mammals' dataset using the alternative preconditioning heuristics to reduce the original points in the dataset from $n$ down to $s$ points. The time it takes for computing the convex hull on the original set of $n$ points is given as $t_n$. The total execution time to compute the convex hull on the reduced set of $s$ points is then given by the time it takes to perform the reduction, $t_r$, plus the time it takes to compute the convex hull on the reduced set, $t_s$. The overall total time is reported here as $t_{rs} = t_r + t_s$; see Fig. 9.

For the TZTM method, the global time for computing the convex hull is shown for 2, 3 and 4 iterations respectively. QuickHull is used as the final convex hull algorithm, in order to have a common factor to compare against.

From an analysis of the results (not shown in Fig. 9), for the AT method, the percentage of time to compute the final convex hull on the reduced set of points is between 10%-30% of the overall time. For the TZTM method, it is of around 10% for two iterations and well under 1% for four iterations. In other words, most of the time is spent in reducing the original number of points.

Interestingly, note that none of these two methods (AT, TZTM) show that reducing the number of points from $n$ down to $s$ gives any advantage for computing the convex hull for this dataset when the reduction time cost is taken into consideration (QuickHull ($n$) is always faster). For Fig. 8, we see that the proposed method for reducing points does give a speedup of at least around three. This is the tradeoff that the proposed method offers: spend time in reducing the points first, and then use a convex hull algorithm to build the final hull and you get an acceleration factor that indeed reduces the overall execution time to build the final hull.

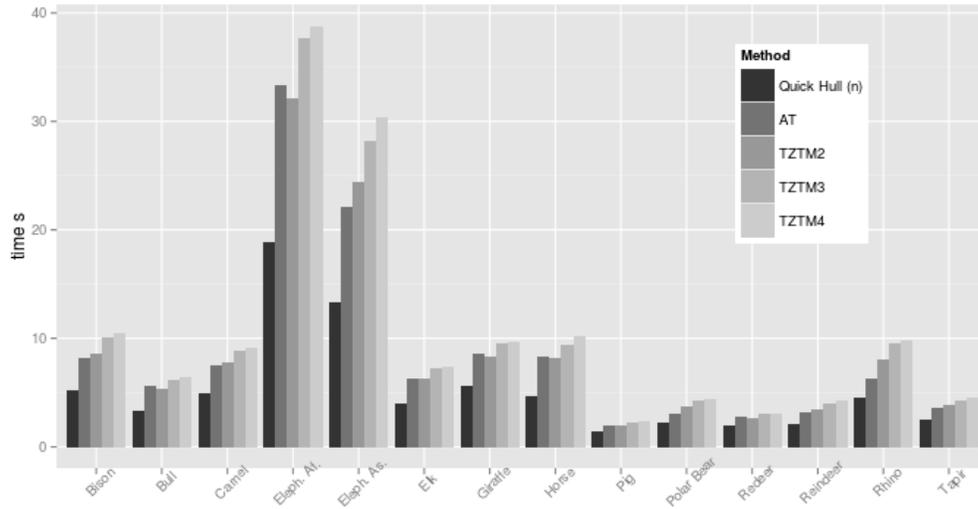

Fig. 9. Execution time spent to compute the convex hull using the two alternative point reduction methods. The bar label QuickHull (n) shows execution time using the original $n$ points. For the AT and TZTM methods QuickHull was used to build the final convex hull on the reduced set of points.

**Proposed heuristic impact on the convex hull computation using Melkman's method**

Reducing points by either AT or TZTM methods does not produce a polyline immediately without some further preprocessing; the method proposed here does as showed earlier in Section 2.1; thus we can use Melkman method directly after the preconditioning.

Fig. 10 shows the execution time when building the convex hull by first reducing $n$ down to $s$ ($t_r$), and then applying Melkman to the reduced set of $s$ points ($t_s$), so the total time is $t_r+t_s$. The figure shows the speed up $t_n/(t_r+t_s)$, where $t_n$ is the time to compute the convex hull on the original set of $n$ points by the QuickHull method; this speed up is of around a factor of three. Thus the proposed method provides a smooth path to apply Melkman's convex hull algorithm to a set of $n$ points without forming a polygonal chain originally; this is actually the case for the points taken from the mammals' dataset. If any other generic method to build a polygonal chain is used, the time to build the polygonal on $n$ points also has to be taken into account. From analysis of the time results it can be calculated that the time to compute the convex hull is under 5% of the overall time, or in other words the time to perform the reduction dominates. Nevertheless, observe that roughly three times faster execution is obtained by building the hull on the reduced set compared to building the hull on the original set. This is similar to the acceleration seen using the QuickHull algorithm.

Empirical evaluation of preconditioning for fast convex hull computation

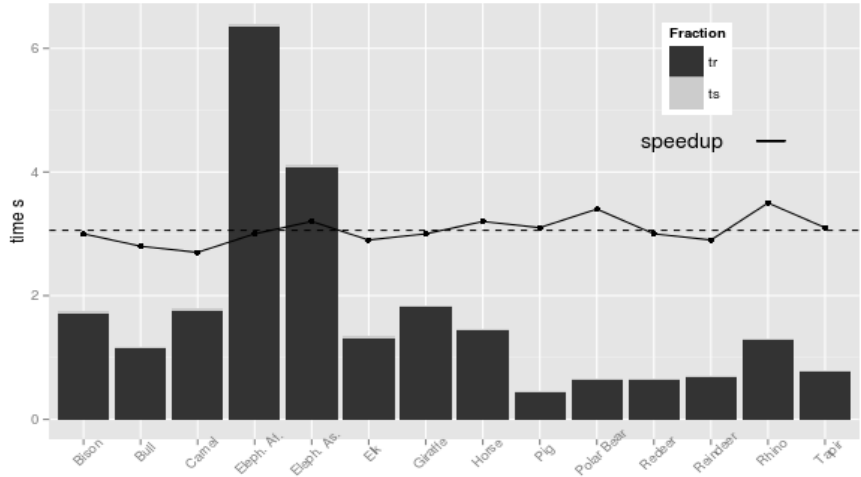

Fig. 10. Execution time to compute the convex hull on the mammals' dataset using Melkman algorithm. Time to perform the reduction from $n$ to $s$ points is $tr$ and time to compute the hull on $s$ points is $ts$. The speedup is plotted against computing the hull on $n$ points with the QuickHull algorithm.

## Discussion

Perhaps a controversial point in our presentation is that we use an array of size $p$ in Algorithm 1 and claim that it can be searched in O($n$) time; we postpone this to the heading "For All Practical Cases" below. For sufficiently dense data sets $p < n$ is expected since the $n$ points are confined into a square of size $p^2$ so for a very dense set $p \approx \sqrt{n}$. For very sparse sets then it may be the case that $p > n$; such sets are not typical of the data characterized by the study here (scanned images). Nevertheless we provide the following analysis for completeness.

Algorithm 1 reduces a set of $n$ points to a set of $s$ points such that $s \leq n$ and $3 \leq s \leq 2p$ valid points; all these valid points are contained in array $L$ as a result. The value $p$ is found by step 1 of the proposed heuristic after all $n$ points had been visited to find the maximum and minimum $x$, $y$ for all points. As this paper restricts $x$, $y$ to be integers, then it is natural to assume the computation to take place using the word RAM model of computation [8]. As such ($x$, $y$) coordinates for points are assumed to be expressed using a computer word size of $w$-bit, so their range is bound by $2^w$, thus $p \leq 2^w$.

During the scan of points it may well be that $L$ contains entries that are not valid points (that is entries in $L$ that keep the original initialized values, representing empty entries in the $y$ direction). Let us refer to these points as gaps in $L$. The purpose of this section is to discuss how to extract, in general, $s$ valid points from an array $L$ of $p$ entries that may have gaps. The relation to the problem size $n$ in cases where $p \leq n$ does not seem to be particularly problematic as the extraction of the $s$ points from $L$ will be bound by $p$ and hence O($n$), a linear case. Now suppose that $p > n$, the extraction of points is still bound by $p$ and then the time complexity is O($p$) instead of O($n$). Below we explain two mechanisms to make the extraction of $s$ points fast in practice. By this we mean that the cost of skipping is very much faster than processing a point and so for all practical purposes the cost of extraction remains close to linear.

**Basic extraction idea**

The basic idea, underlying the two mechanisms to be discussed, is to think of $L$ as composed of blocks of $w$ entries each. A word of $w$-bit is then associated to each block, where individual bits in the word are used to represent whether an entry in a block of $L$ contains at least a valid point. For example $L = [(1,4), (2, 4), (2,5), (3,3), (2,3)]$ corresponding to the small example of Fig. 1 (center) can be represented as a single block of 5 entries and as such a single 5-bit word is associated with it. Notice that in this example, all 5 points are valid and then a 5-bit word with "11111" can be used to denote this situation. Suppose, instead, that points in columns $x = 2$ and $x = 4$ are removed then $L = [(1,4), (6, -1), (2,5), (6,-1), (2,3)]$ and the points in $L$ can be represented with the 5-bit word "10101" to indicate which entries have valid data (with the gaps denoted by a zero bit).

The extraction idea capitalizes on the fact that there exist fast bit manipulation routines to extract set bits in a $w$-bit word. Current machines are built with word sizes of either $w = 32$ or $w = 64$ bits. Common bit manipulation routines for extracting set bits are the so called *count trailing zeros* ("ctz") or *count leading zeros* ("clz") functions. For a $w$-bit machine they take constant time or O(1). For example, clz("**1**0101") = 4 indicating that the leftmost set bit position is position 4 (in bold, counted from the right as bit 0). Similarly, ctz("1010**1**") = 0 indicating that bit at position 0 is set. The procedure in Table I can be used to extract all the 3 set positions (*pos*) within the 5-bit word "10101" (*x*) using the "ctz" function with an update step involving the current value of the word $x$ XOR (^) with a left shifted version of the position (1 << *pos*) found for the current word value $x$ [13]. Other variations for the update step are also used in practice such as $x = x$ AND $\sim(1 <<$ *pos*$)$ with $\sim$ indicating the unary NOT operation on bits [14].

Table I. Extracting the set bit positions (*pos*) for a 5-bit word *x* using the counting leading zeros function ctz.

| Iter. | $x$ | pos = ctz($x$) | $x = x$ XOR $(1 <<$ pos$)$ | *Pseudo Code* |
|---|---|---|---|---|
| 0 | 10101 | 0 = ctz(10101) | 10100 | **while** x != 0 **do**: |
| 1 | 10100 | 2 = ctz(10100) | 10000 | pos = ctz(x) |
| 3 | 10000 | 4 = ctz(10000) | 00000 | x = x ^ (1 << pos) |

Observe that this extraction method can be captured as a simple while loop. The while loop goes through as many iterations (3 iterations in this example) as set bits are within a $w$-bit word for which there is a fast *ctz* function implementation. Most standard machines built today (including embedded microprocessors) do support fast "ctz" ("clz") functions.

**Extraction using an array of bits**

Define a bit array $M[i]$ $i = 1, …, p$ with all bits initialized to 0. Now, set $M[x_i] = 1$ for each $L[x_i] = (ly, hy)$. At the end of algorithm 1 the bit array $M$ will contain zeros at $i$ positions where no valid points are registered in output array $L$. Since the number of gaps is the amount $(p - s)$ and the number of sets bits is $s$, the while loop is executed in O($s$) time which in the worst case is O($n$) when $s = n$. It follows that given Algorithm 1, in principle we can extract the bits into an ordered list in O($n$) time although we need O($p$) space for $L$.

On desktop processors `__builtin_ctz`, `__builtin_ctzl` are "ctz" functions for $w = 32$ and $w = 64$ respectively [15]; these are supported as native hardware instructions in their instruction dataset. Similarly, `__builtin_clz` and `__builtin_clzl` are "clz" functions for $w = 32$ and $w = 64$ respectively. Consequently for $p \leq w$, $M$ can be reduced to a single word.

More generally consider when $p > w$ then in a word RAM model we can only operate on $w$-bit words in O(1) time. Suppose $p$ is bounded by $2^w$. A trivial but effective approach is to partition the $M$ array into $w$-bit word blocks. Any $i$, $i = 1, …, p$ would correspond to word $(i-1) >> \log_2 w$ block and to bit position $(i-1)$ & $(w-1)$ within that block word. For example for $w = 32$ bits, $i = 103223$ would correspond to block 3225 and to bit position 22 within that block word, so that $3225*32 + 22 + 1 = 103223$. This applies in general for $p \leq 2^w$, block $b = (i-1) >> \log_2 w$ and the position within the block $pb = (i-1)$ & $(w-1)$. Thus, repeating the example for a $p$ value requiring only 24 bits, gives block $b = 3225$ and $pb = 22$, so that $i = b*w + pb + 1$. Note that for a $p$ value expressed in 24 bits $M$ is a bit array of $2^{24}$ bits arranged as 524288 block words of 32-bit each; operations are now performed on 32-bit unsigned integers for a fast set bit extraction mechanism within each integer as shown in Table I.

**Extraction using a tree structure**

The trivial mechanism here suggested for preconditioning data skips over gaps with block words but in a sparse situation the problem of skipping gaps is just shifted to a block level. In other words, it does not skip complete empty block words. It seems natural that an approach to skip whole block words with gaps demands extending that suggestion to different levels. Thus, we can associate each $w$-bit block word to a single bit and set this bit (or not) depending on whether (or not) the word is non-zero. A group of $w$ block words is then mapped to yet another $w$-bit block word. This can be organized as a tree structure.

A van Emde Boas tree [16] to provide this functionality seems a good choice for two reasons. Firstly, some ordering is implicit in the tree and so no explicit sorting is required, and secondly operations on the tree can be made fast. Tree operations on a universe $u$, $u = p$ for the problem in this paper, can be performed in time complexity O($\log \log p$); if $p$ can be written as 24 bits, this is 5 steps.

A close examination for the problem of extracting the reduced set of $s$ points from the output array $L$ of Algorithm 1 reveals that there is no need to implement a full van Emde Boas tree. There are various reasons for this. The leaves of the tree should be pruned to be composed of $w$-bit words to exploit the basic extraction ideas previously suggested; nodes can be arranged as arrays of $w$-bit words. Building $M$ requires only insert operations. Recovering all the ones in $M$ requires a successor operation which in turn requires a *min* operation on nodes of the tree. It happens that the function "ctz" can be used for a fast *min* operation on nodes or leaves. So, we can use a $w$-ary tree.

An analysis shows that a $w$-ary tree would have a height $h$ given by $h = \log_b((w-1)*p + w)-2$. For $p = 2^{24}$, this is $h = 3$ using 64-bit words as nodes. This is directly related to how you would partition 24 bits into sections of 6-bits each ($6 = \log_2 64$). An analysis also shows that the overall tree has a cost of $m$ $w$-bit words with $m = (1/(w-1))*(w^{h+1}-1)$. For $p$ expressed in 24 bits, this is 266305 64-bit words while a single $M$ array as the one suggested would have $2^{18} = 262144$ 64-bit words, or an overhead of 4161 64-bit words, this is roughly 1.5% more memory space in this case. Thus, the tree does not seem particularly costly. Using this tree structure to extract *s* valid points recorded in an array *L* of $p = 2^z$ entries requires $h+1$ steps to either insert or find a particular position *i*, $i = 1, …, p$. The time to visit every node in the tree is $(p-1)/(w-1)$ and as each valid set bit in each node has to be visited at least once to account for all valid points in *L*, the maximum time to access all combinations of valid points is $w(p-1)/(w-1)$. As $w/(w-1)$ approaches one as *w* approaches infinity, then theoretically the whole mechanism to extract valid points is $O(p)$.

The tree however has a capability to skip through complete block words of gaps. Since the distribution of empty blocks depends on the data we evaluate this empirically using realistic datasets. Fig. 11 shows the execution time of two mechanisms to extract set bits from an array of $p = 2^{24}$ bits. The density of bits set in the array *s* to the array size *p* expressed as *s/p* is varied from 5% to 85%. The preconditioning mechanism for reducing data used for the results of this paper are implemented as an Array, while now are implemented as a Tree for comparison. The figure gives the time to do *s* inserts (for setting *s* bits) and the time to scan all the *s* set bits and actually extracting them. The total time of inserts plus a scan is presented for easier inspection. A regression analysis on these data shows a linear trend in execution time of $1.23 \times 10^{-6} n$ seconds and of $2.64 \times 10^{-6} n$ seconds for the array and tree respectively. The figure confirms that inserting into the array is faster than into the tree. Also that extracting the bits may be faster by using the tree than the array, although overall the array is twice as fast as the tree. There is evidence in the figure to suggest that the trivial mechanism of a simple array is also practical.

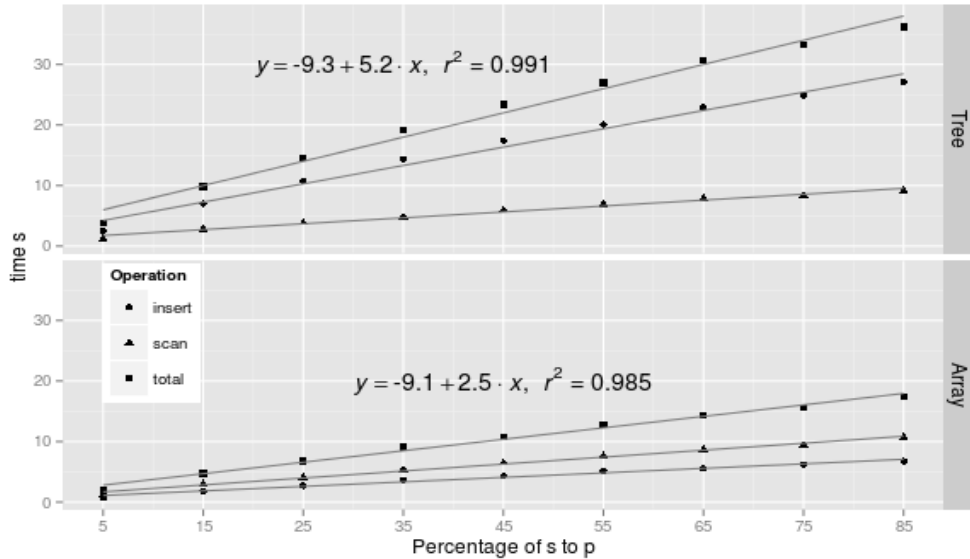

Fig. 11. Given a bit vector M of $p = 2^{24}$ bit entries where *s* bits are set to 1, this graph shows the time to extract the *s* set bit positions from M using either a Tree data structure or a simple Array.

**For all practical purposes**

Regarding complexity there are some nuances to our argument which require further elaboration. We now elaborate on our claim that the method is $O(n)$.

At the beginning of this section it was stated that by the usual standards of complexity analysis it can be argued that the method is O($p$) in space and time complexity. Cases where $p \leq n$ do not seem to be particularly problematic as the extraction of the $s$ points from $L$ will be bound by $p$ and hence O($n$), a linear case. When the data becomes increasingly sparse then the method is challenged on two fronts. First, that storing the points is O($n$) but $p > n$. Second, that we will encounter significant gaps in the storage array. The extraction technique shows that the time here is related to the number of words used for the bit encoding and in particular the number of empty words. Because an array of $p$ bits can be organized as $p/w$ words of $w$-bit each, then it follows that, in the worst case, for $p > n$, $p \leq n + k$ where $k$ is the number of $w$-bit words without a single bit set on it or empty words; clearly the method is O($p/w$). As skipping an empty word has the same cost as extracting a valid bit, then you want $k < n$; as $k \leq (p-n)/w$ then $(p-n)/w < n$ implies that the method remains O($n$) as long as a condition for $p < n(w+1)$ holds. For instance, say $n = 19$ and $p = 64$ organized as 8 words of 8-bit each; if $p < 171$ (19*9) then $k = (171-19)/8 = 19$ otherwise skipping empty words would dominate. This is related to the density of the data set or the relationship between the bounding box containing the hull and the number of points. From Fig. 4 we see that for the dataset in this paper the condition for $p$ above holds for $n > 72$ points on average or densities as small as 0.002%. For extremely low densities, where $p < n(w+1)$ does not hold, it probably is not worth trying to reduce any points at all by any method; thus for all practical cases our proposed method behaves O($n$).

## Conclusions

- The proposed method for reducing a set of $n$ points down to a set of $s$ points achieves over 98% of reduction for the dataset used in this paper. It gave at least an extra 7% better reduction than the AT method. The TZTM method needs at least an extra iteration of the AT method to achieve a better reduction than the method proposed here (Fig. 3). However, the total time to achieve the reduction by the method proposed here showed faster execution time than AT and TZTM (Fig. 5).
- The proposed method shows a percentage in the reduction of points that was under 2% from the best possible theoretical percentage of reduction of 1-2/$p$ (Fig. 4). Although a further reduction is also possible (running Algorithm 1 one more time) the overall advantage, when the extra execution time is taken into account, is not significant and thus the suggestion is to keep a single pass of Algorithm 1 (Fig. 7).
- The proposed method differs from many methods in the sense that it can be thought of as a process of including points that are candidates to belong to the final convex hull rather than a process of discarding points that cannot belong to the hull.
- The proposed method has a preliminary step for finding a bound for the output array size ($L$) that contains the reduced set of points. This step is of around 30% of the total time to complete the reduction. This compares to a similar step for the AT method to find some extreme points on the original set that accounts for around 20% of the total time to complete a reduction using AT's method. The proposed method is around four times faster to make a reduction than AT method for the dataset used in this paper (Fig. 6).
- The proposed method provides an overall acceleration to compute the convex hull for a dataset of $n$ points by first reducing the points from $n$ down to $s$ and then computing the convex hull on the reduced set of $s$ points rather than computing the convex hull on the original set of $n$ points with the same convex hull algorithm. This was shown consistently using convex hull of different time complexities namely QuickHull, Graham Scan, Jarvis March (Fig. 8). This acceleration was seen to be often more than a factor of 2 and indeed over a factor of 30 for some instances of the dataset. This work did not find that either the AT or the TZTM method provides an overall acceleration in the same way; this is observed for the dataset used in this paper (Fig. 9).

- The reduced set of points, $s$, proposed by the method in this paper, is recovered in such a way that a simple polygonal chain can be formed from these $s$ points. This allows the linear convex algorithm known as Melkman to be applied in a straightforward manner. For Melkman's algorithm, the method also shows acceleration for computing the overall convex hull. An acceleration factor of around three is shown for the dataset in this paper (Fig. 10). The reduced set of points obtained by the AT and TZTM methods do not prepare the points directly for Melkman's convex hull algorithm.
- Two mechanisms for fast extraction of valid points from the output array $L$ containing the reduced set of points were empirically evaluated with the final suggestion that a simple linear array composed of $w$-bit words is a good choice (Fig. 11). The mechanism exploits common bit manipulation routines available from commercial processors assuming the word RAM model of computation.
- Although the method works for data with integer coordinates only, a suitable scaling can be used to convert data in floating point coordinates. In fact, the original 3D dataset used in this study has floating point coordinates.

This paper has presented a method to reduce the number of $n$ points with integer coordinates $(x, y)$ to a smaller set of $s$ points composed of three easy computational steps. The reduced set of $s$ points is contained in an array $L$ of $p$ entries, with $p$ related to $min$, $max$ values of one of the $x$, $y$ coordinates. Array $L$ however, can have entries that do not correspond to any original points and so must be skipped. These entries are referred to as gaps. Although it is guaranteed that the method will always produce $s \leq n$ points, extracting valid points from array $L$ has $p$ as an upper bound and so O($p$) for cases where $n < p$. We further argue that if $p < n(w+1)$ the method remains O($n$). The extraction of valid points from $L$ is O($n$) for cases where $n \geq p$; this condition is satisfied by all the instances of the dataset examined in this paper. The paper suggests and also evaluates, in an empirical manner, two different mechanisms for fast extraction of points, and finds that even when $n < p$ the two mechanisms show a linear trend for practical cases of different densities of gaps to valid points within a bit array. Authors are validating a method to build the final convex hull based from the reduced set contained in array $L$ without an explicit need to call an external known convex algorithm. As it currently stands, the array $L$ can be given to any convex hull algorithm of any time complexity to build a final convex hull on $s$ points that is the same hull as the one obtained from the original set of $n$ points providing an extra acceleration of at least four times when compared to previous existing preconditioning methods.

# Appendix

*Correctness of Algorithm 1:* The claim is that the convex hull of the $n$ original points contained in a box of size $p \times p$ points is the same as the convex hull obtained from the smaller set of $s$ points contained in the output array $L$ after the $n$ points are processed by Algorithm 1. Reasoning by induction towards asserting this claim follows like this. The smallest convex hull is a triangle. This triangle has to be represented by three distinct points in array $L$, $s = 3$ is the basis. Either case 1, three entries with valid points are found at positions $i \neq k \neq l$, or case 2, two points at position $i$ and one point at position $k$, with $i \neq k$, for any $1 \leq i, j, k \leq p$. The original set of points may have been composed of $n \geq s$ points. In case 1, clearly $n = 3$, so no reduction was achieved. In case 2, an entry at position $i$ of $L$ had two valid points, so, $ly_i \neq hy_i$ and given that $1 \leq ly_i, hy_i \geq q$ implies that there could have been collinear points in between the two extreme valid points $(i, ly)$ and $(i, hy)$, so $n \leq q$-2 for the bounding box at column $i$ and $n \leq q - 1$ in general, so Algorithm 1 will reduce points for any $q \geq 4$. By definition, a collinear point in between an edge of a convex hull cannot be part of the hull, so the claim holds. Assume, $L$ contains $s$ valid points, the claim holds in general if those $s$ points captures the same convex hull of $n > s$ points. Thus, consider the effect on array $L$ with $s$ valid points by adding an extra point within the limits of the bounding box. Three cases are distinguished. Case 1, the new point, is added at box column $i$ where no valid point is previously found in $L$. Case 2, the new point $(x, y)$ is added at box column $i$ where one valid point is previously found in $L$. Case 3, the new point is added at box column $i$ where two valid points are previously found in $L$. For case 1, algorithm 1 does not make any further assumption on whether the point might be removed from the final convex hull and so a point $(x, y)$ gets inserted into $L[x]$ as $(y, y)$. This situation is promoted to case 2 of the current reasoning if yet an extra point is added. For case 2, $L[i]$ had a valid point as $(ly, ly)$ [or $(hy, hy)$ as $ly = hy$]. The new point $(x, y)$ gets inserted into $L[x]$ either as $(ly, y)$ if $y > ly$, or as $(y, hy)$ if $y < ly$ ($ly = hy$) or remains as $(ly, ly)$ if $y = ly$. Observe, in the latter possibility $n > s$. This case is then promoted to case 3 if yet an extra point is added (or remains as case 2). For case 3, either the new point $(x, y)$ replaces an existing valid point in $L[x]$ as minimum or maximum or it is discarded by being collinear to $(i, ly)$ and $(i, hy)$. In any case $n > s$ ($n = s + 1$) and so the claim holds.